\documentclass{article}

\usepackage{microtype}
\usepackage{graphicx}
\usepackage{subfigure}
\usepackage{booktabs} 

\usepackage{hyperref}

\usepackage[noend]{algorithmic}

\usepackage[accepted]{mlsys2025}

\usepackage{stfloats}

\usepackage{multirow} 
\usepackage{booktabs}
\usepackage{float}
\usepackage{tikz}
\usepackage{amsmath}
\usepackage{ifthen}
\usepackage{mfirstuc}

\usepackage[skip=0.1in]{caption}
\begin{document}

\newcommand{\blackcirclednumber}[1]{%
    \tikz[baseline=(char.base)]{%
        \node[shape=circle, fill=black, inner sep=0.2pt, 
              text=white, font=\sffamily\footnotesize] (char) {#1};
    }%
}

\newif\ifusewordscircle
\newif\ifuseemptycircle

\usewordscirclefalse
\useemptycirclefalse

\newcommand{\circlednumber}[1]{
  \ifuseemptycircle
  \else
    \ifusewordscircle
      \ifcase #1
        \or \textbf{Firstly,}\or \textbf{Secondly,}\or \textbf{Thirdly,}
        \else \textbf{Fourthly,} 
      \fi
    \else
      \blackcirclednumber{#1}
    \fi
  \fi
}

\newcommand{\usewordcircle}{
  \usewordscircletrue
  \useemptycirclefalse
}
\newcommand{\usenumbercircle}{
  \usewordscirclefalse
  \useemptycirclefalse
}
\newcommand{\useemptycircle}{
  \usewordscirclefalse
  \useemptycircletrue
}

\newcommand{\maxGoodputImprovementMMEAcrossModelsBaselines}{1.3}
\newcommand{\maxGoodputImprovementPOPEAcrossModelsBaselines}{1.6}
\newcommand{\maxGoodputImprovementTextCapsAcrossModelsBaselines}{3.7}
\newcommand{\maxGoodputImprovementTextVQAAcrossModelsBaselines}{2.3}
\newcommand{\maxGoodputImprovementVizWizAcrossModelsBaselines}{2.1}
\newcommand{\maxGoodputImprovementAcrossDatasetsModelsBaselines}{3.7}
\newcommand{\minGoodputImprovementQwentwoVLsevenBvLLMAcrossDatasets}{1.3}
\newcommand{\maxGoodputImprovementQwentwoVLsevenBvLLMAcrossDatasets}{2.4}
\newcommand{\minGoodputImprovementQwentwoVLsevenBSGLangAcrossDatasets}{2.0}
\newcommand{\maxGoodputImprovementQwentwoVLsevenBSGLangAcrossDatasets}{7.9}
\newcommand{\minGoodputImprovementLLaVAonepointfivesevenBvLLMAcrossDatasets}{1.3}
\newcommand{\maxGoodputImprovementLLaVAonepointfivesevenBvLLMAcrossDatasets}{3.7}
\newcommand{\minGoodputImprovementLLaVANeXTsevenBvLLMAcrossDatasets}{1.2}
\newcommand{\maxGoodputImprovementLLaVANeXTsevenBvLLMAcrossDatasets}{2.1}

\newcommand{\goodputours}{3.9}
\newcommand{\goodputourswohybridEPD}{2.6}
\newcommand{\goodputourswohybridEPDandsched}{2.2}

\newcommand{\systemname}{HydraInfer}
\newcommand{\titlename}{\systemname{}: Hybrid Disaggregated Scheduling for Multimodal Large Language Model Serving}

\newcommand{\keywordsname}{MLLM Inference System, Request Scheduling, Encode-Prefill-Decode Disaggregation}

\newcommand{\formatSystemname}[1]{\texttt{#1}}

\newcommand{\formatDisaggregationMethod}[1]{\textit{#1}}

\newcommand{\formatInstanceType}[1]{\textit{#1}}

\usewordcircle

\twocolumn[
\mlsystitle{\titlename{}}
\mlsyssetsymbol{equal}{*}
\begin{mlsysauthorlist}
\mlsysauthor{Xianzhe Dong}{ustc}
\mlsysauthor{Tongxuan Liu}{ustc}
\mlsysauthor{Yuting Zeng}{ustc}
\mlsysauthor{Xiaoyang Zhao}{jd}
\mlsysauthor{Weizhe Huang}{jd}
\mlsysauthor{Liangyu Liu}{ustc}
\mlsysauthor{Yang Liu}{ustc}
\mlsysauthor{Siyu Wu}{buaa}
\mlsysauthor{Yu Wu}{ustc}
\mlsysauthor{Hailong Yang}{buaa}
\mlsysauthor{Ke Zhang}{jd}
\mlsysauthor{Jing Li}{ustc}
\end{mlsysauthorlist}
\mlsysaffiliation{ustc}{University of Science and Technology of China}
\mlsysaffiliation{buaa}{Beihang University}
\mlsysaffiliation{jd}{JD.com}
\mlsyscorrespondingauthor{Tongxuan Liu}{tongxuan.ltx@mail.ustc.edu.cn}

\mlsyskeywords{\keywordsname{}}
\vskip 0.3in

\begin{abstract}
Existing MLLM inference systems are typically designed based on the architecture of language models, coupling image processing and language processing.
This design struggles to accommodate the heterogeneous demands of different stages in terms of computational resources, memory access patterns, and service-level objectives (SLOs), leading to low resource utilization and high request latency, ultimately failing to meet the service requirements of diverse inference scenarios.
To address these challenges, we propose \formatSystemname{\systemname{}}, an efficient MLLM inference system that adopts a Hybrid Encode-Prefill-Decode (EPD) Disaggregation architecture.
By scheduling the three stages — encode, prefill, and decode — onto separate heterogeneous inference instances, the system flexibly reallocates resources across stages, significantly reducing idle computation, alleviating resource bottlenecks, and improving overall system throughput and scalability.
In addition, \formatSystemname{\systemname{}} supports a stage-level batching strategy that enhances load balancing, enables parallel execution of visual and language models, and further optimizes inference performance. 
Experiments under real multimodal inference workloads demonstrate that \formatSystemname{\systemname{}} can achieve up to \maxGoodputImprovementAcrossDatasetsModelsBaselines{}× higher inference throughput compared to state-of-the-art systems (e.g., vLLM, SGLang) while meeting the 90th percentile request SLO.
\end{abstract}
]

\printAffiliationsAndNotice{} 

\section{Introduction}
\label{intro}
In recent years, Multimodal Large Language Models (MLLMs) ~\cite{qwenvl,li2023blip2bootstrappinglanguageimagepretraining,llava1.5,team2023gemini,zhu2023minigpt4enhancingvisionlanguageunderstanding,hu2024minicpm,lu2024deepseek,qwen2vl, bai2025qwen2} have demonstrated impressive capabilities across various domains, including image understanding and visual question answering. 
The typical inference process of MLLMs typically consists of three stages: the image encode stage extracts visual features from input images; the prefill stage feeds visual features and the text prompt into the language model to generate the first output token; and the decode stage iteratively generates subsequent tokens based on the cache~\cite{Yin_2024}. 
Notably, input images are converted into hundreds or even thousands of tokens during inference~\cite{cai2024vipllavamakinglargemultimodal}, requiring significantly more GPU resources compared to large language models~\cite{chen2024imageworth12tokens}. Consequently, reducing the inference cost of MLLMs has emerged as a critical research topic.

There are now many studies aimed at improving the performance of multimodal inference systems.
Firstly, current systems such as vLLM~\cite{vllm}, SGLang~\cite{sglang}, TGI~\cite{text-generation-inference} performs vision and language processing sequentially, failing to exploit inherent cross-modal parallelism. This results in suboptimal hardware utilization. Secondly, although techniques such as continuous batching~\cite{orca}, chunked prefill and stall-free scheduling~\cite{sarathi-serve} optimize language model scheduling for throughput and latency, they prevent stage-specific optimization crucial for multimodal inference. As shown in Figure~\ref{fig:batchsize_analysis}, different stages have different characteristics. Consequently, these methods struggle to trade off throughput and latency, frequently violating Time-between-Tokens (TBT) Service Level Objectives (SLOs). Thirdly, current large-scale inference systems deploy disaggregated architecture across multiple inference instances ~\cite{distserve, singh2024efficiently, qiu2025towards, guo2025rserveoverlappingencodingprefill}. Although disaggregation methods such as \formatDisaggregationMethod{E+P+D} (encode, prefill, decode all separated), \formatDisaggregationMethod{EP+D} (encode, prefill co-located, separated with decode), and \formatDisaggregationMethod{ED+P} (encode, decode co-located, separated with prefill) have specific advantages, existing systems use a single and fixed disaggregation method regardless of varying scenarios, resulting in low goodput. 

To address these limitations, we introduce \formatSystemname{\systemname{}}, a novel MLLM inference system designed and implemented from the ground up. 
Firstly, we propose a dual-stream architecture that allocates image encoding tasks to the vision stream while processing prefill and decode tasks in the language stream. This enables parallel execution of heterogeneous stages across requests, significantly improving hardware utilization and system throughput.
Secondly, we propose a Stage-level Schedule strategy, which decompose each request into fine-grained stages. By applying stage-specific scheduling and batching optimizations, this strategy eliminates bottlenecks inherent in coarse-grained scheduling while enabling precise execution time control.
Thirdly, we propose a Hybrid Encode-Prefill-Decode disaggregation architecture, which enables each instance to execute configurable stage subsets with automatically adjust the appropriate disaggregation method based on historical trace, maximizing the system's goodput.

To validate the effectiveness of \formatSystemname{\systemname{}} , we conduct comprehensive evaluations across various MLLMs and representative tasks, including image captioning ~\cite{textcap}, visual question answering ~\cite{vizwizvqa, textvqa}, model evaluation~\cite{pope, mme}. The results of experiments demonstrate that \formatSystemname{\systemname{}} consistently outperforms state-of-the-art inference frameworks including vLLM and SGLang.
Specifically, under different SLO constraints, \formatSystemname{\systemname{}} achieving up to 
\maxGoodputImprovementMMEAcrossModelsBaselines{}x, 
\maxGoodputImprovementPOPEAcrossModelsBaselines{}x, 
\maxGoodputImprovementTextCapsAcrossModelsBaselines{}x, 
\maxGoodputImprovementTextVQAAcrossModelsBaselines{}x, 
and \maxGoodputImprovementVizWizAcrossModelsBaselines{}x 
goodput improvements on MME~\cite{mme}, POPE~\cite{pope}, TextCaps~\cite{textcap}, TextVQA~\cite{textvqa}, and VizWiz ~\cite{vizwizvqa} datasets, respectively.
Our main contributions are as follows:
\begin{itemize}
\item We identify and analyze key challenges in existing inference systems for MLLMs, particularly in parallelism, scheduling granularity, and disaggregation method selection.
\item We design Hybrid Encode-Prefill-Decode disaggregation architecture, and automatically selects the optimal disaggregation method based on workload and SLO profiles for flexible resource allocation.
\item We propose a stage-level scheduling strategy, which applies batching optimizations at different inference stages and incorporates a multi-stream parallel execution model to improve system throughput and resource efficiency.
\item Comprehensive evaluation across diverse models and scenarios demonstrating strong generality and significant service capacity improvements, substantially advancing the state of multimodal inference systems.
\end{itemize}
\section{Background}

\subsection{MLLM Autoregressive Inference}

\begin{figure}[t]
  \centering
  \includegraphics[width=\linewidth]{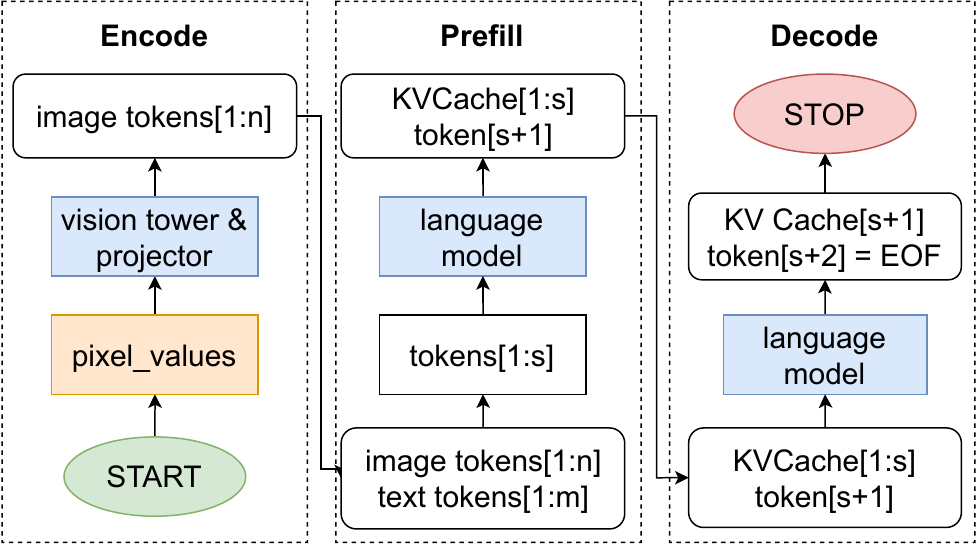}
  \caption{Vision-language model autoregressive inference process.}
  \label{fig:autoregression}
\end{figure}
As shown in Figure~\ref{fig:autoregression}, the inference process of a transformer-based~\cite{vaswani2017attention} vision-language model can be divided into three distinct stages~\cite{Yin_2024}.
The first stage is the encode stage, where the model's vision tower and projector process the image of request to obtain the image tokens $I[a_1, a_2, …,a_n]$.
The second stage is the prefill stage. During this stage, the request prompt is encoded into text tokens $T[b_1, b_2, …,b_m]$, and then are concatenated with image tokens to form the final input $X[x_1, x_2, …,x_s]$, where $s = n + m$. This input $X$ is fed into the large language model to generate a token $x_{s+1}$ and a series of $kvcache$ used for the decode stage.
The third stage is the decode stage to generate tokens iteratively until the generated token is $\textless$eos$\textgreater$ or the maximum token limit is reached. Due to the data dependencies in the decode stage, this stage is executed sequentially.

\subsection{Inference Engine Performance Metrics}
\label{sec:background_metrics}
\textbf{TTFT (Time-to-First-Token)} measures the time from when a request arrives at the system until the first output token is produced, reflecting the system's responsiveness.
\textbf{TBT (Time-between-Tokens)} measures the interval between consecutive output tokens for a request, indicating the overall smoothness of the response.
\textbf{Throughput} refers to the number of requests processed per second.
Requests are associated with \textbf{TTFT SLOs} and \textbf{TBT SLOs}. If a request's TTFT is less than its TTFT SLO and 90\% of its TBT values are below the TBT SLO, the request is considered to have met the \textbf{SLO (Service Level Objective)}.
\textbf{SLO attainment} is defined as the percentage of requests that meet their SLOs out of all requests.
\textbf{Goodput} is defined as the maximum request rate at which the SLO attainment reaches at least 90\%.
Our optimization objective is to maximize the per-GPU goodput.

\section{Motivation}
\subsection{Insufficient Parallelism}
\label{sec:motivation_parallel_vision_language_model}

\begin{table}[t]
\caption{Symbols related to the architecture of the model and the batch of requests.}
\centering

\begin{tabular}{ll}
\toprule
Notation & Description \\
\midrule
$B$ & The number of batched requests \\
$S$ & The length of prompt \\
$T$ & The number of tokens per image \\
$H$ & Input dimension of the hidden layer \\
$M$ & The number of attention heads \\
\bottomrule
\end{tabular}

\label{tab:latency_model_notation}
\end{table}
\begin{table}[t]
\centering
\caption{Arithmetic intensity and Memory Acceess of primary operations in MLLMs.}
\resizebox{\linewidth}{!}{
\begin{tabular}{llccc}
\toprule
Operation & E/P/D & FLOPS & Memory Access \\
\midrule
\multirow{3}{*}{QKVO Proj.} & encode & $8BTH^2$ & $8BTH+4H^2$ \\
 & prefill & $8BSH^2$ & $8BSH+4H^2$ \\
 & decode & $8BH$ & $8BH+4H^2$ \\
\hline
\multirow{3}{*}{FFN} & encode & $16BTH^2$ & $10BTH+8H^2$ \\
 & prefill & $16BSH^2$ & $10BSH+8H^2$ \\
 & decode & $16BH^2$ & $10BH+8H^2$ \\
\hline
\multirow{3}{*}{Attention} & encode & $4BT^2H$ & $4BTH+2BT^2M$ \\
 & prefill & $4BS^2H$ & $4BSH+2BS^2M$ \\
 & decode & $4BSH$ & $4BSM+2BH(S+1)$ \\
\bottomrule
\end{tabular}
}
\label{tab:latency_model}
\end{table}

\begin{figure}[t]
  \centering
  \includegraphics[width=\linewidth]{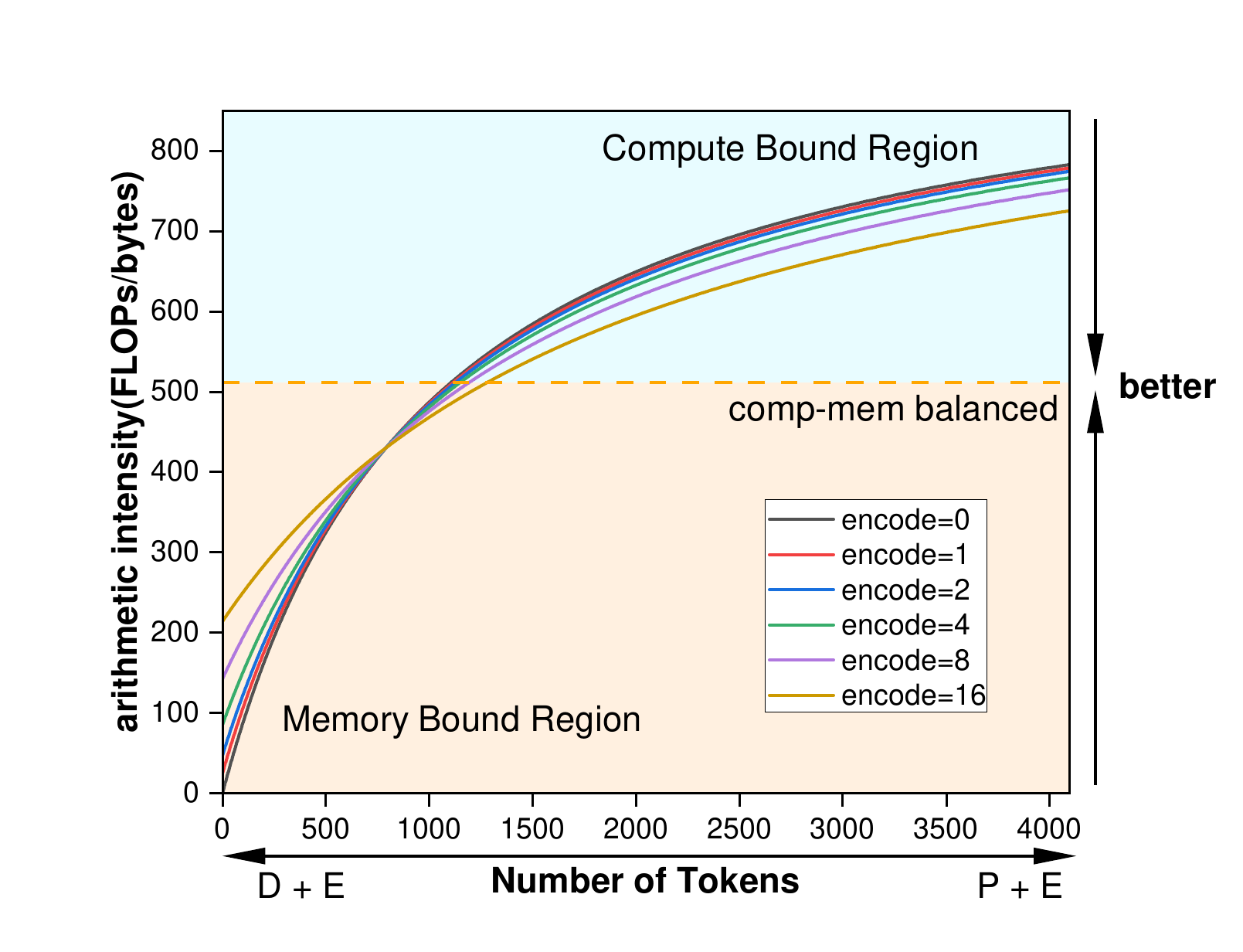}
  \caption{Arithmetic intensity trend for LLaMA1.5-7B linear operations with different number of images and tokens. }
  \label{figs:latency_model}
\end{figure}
\textbf{Diverse Compute and Memory Characteristics.} The prefill stage is primarily compute-bound, while the decode stage is memory-bound. To optimize resource utilization, it is crucial to balance compute-intensive and memory-intensive operations~\cite{sarathi-serve}. 
Our theoretical analysis of FLOPs and memory access patterns for key operations in MLLMs, detailed in Appendix~\ref{sec:latency_model}, with the results summarized in Tables~\ref{tab:latency_model_notation} and~\ref{tab:latency_model}, shows that the encode stage has characteristics that lie between those of the prefill and decode stages in terms of computational and memory usage. Figure~\ref{figs:latency_model}  illustrates the relationship between arithmetic intensity and token count under varying image batch sizes. When the token count is small (i.e., during the decode stage), the operations are memory-bound; in this region, increasing the number of images in the batch raises arithmetic intensity. Conversely, when the token count is large (i.e., during prefill stage), the workload is compute-bound, and batching encode with prefill reduces arithmetic intensity. 

\textbf{Takeaway-1:} The distinct compute and memory characteristics of each stage suggest the potential for optimization through parallel execution of vision and language models.

\subsection{Suboptimal Scheduling Granularity}
\begin{figure}[t]
  \centering
  \includegraphics[width=0.9\linewidth]{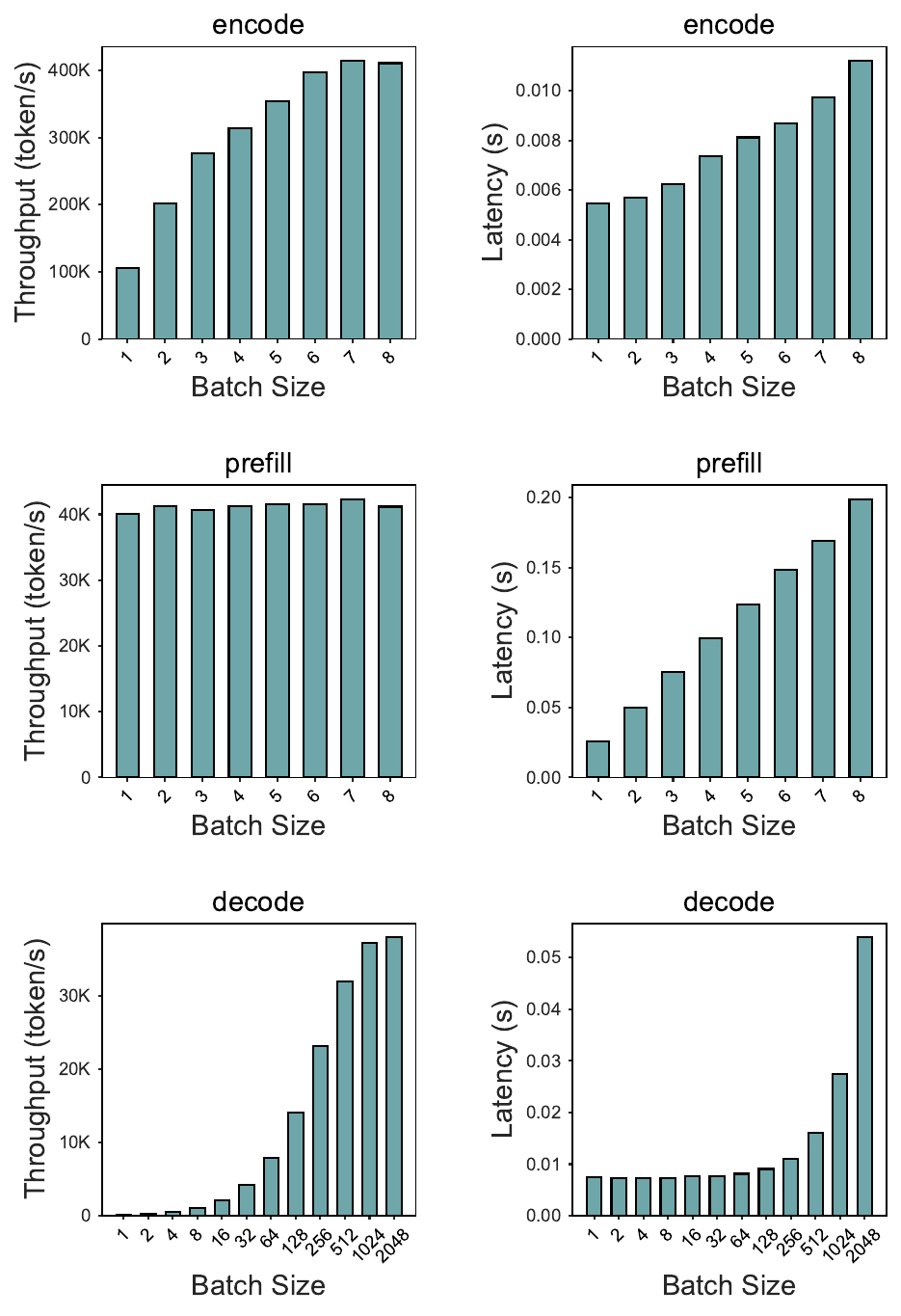}
  \caption{The performance of the LLaVA-1.5-7B model under varying batch sizes. For both the prefill and decode stages, we use a prompt length of 1024 tokens. For the encode stage, the input image resolution is 336×336, corresponding to 576 visual tokens per image.}
  \label{fig:batchsize_analysis}
\end{figure}

\begin{figure}[t]
  \centering
  \includegraphics[width=\linewidth]{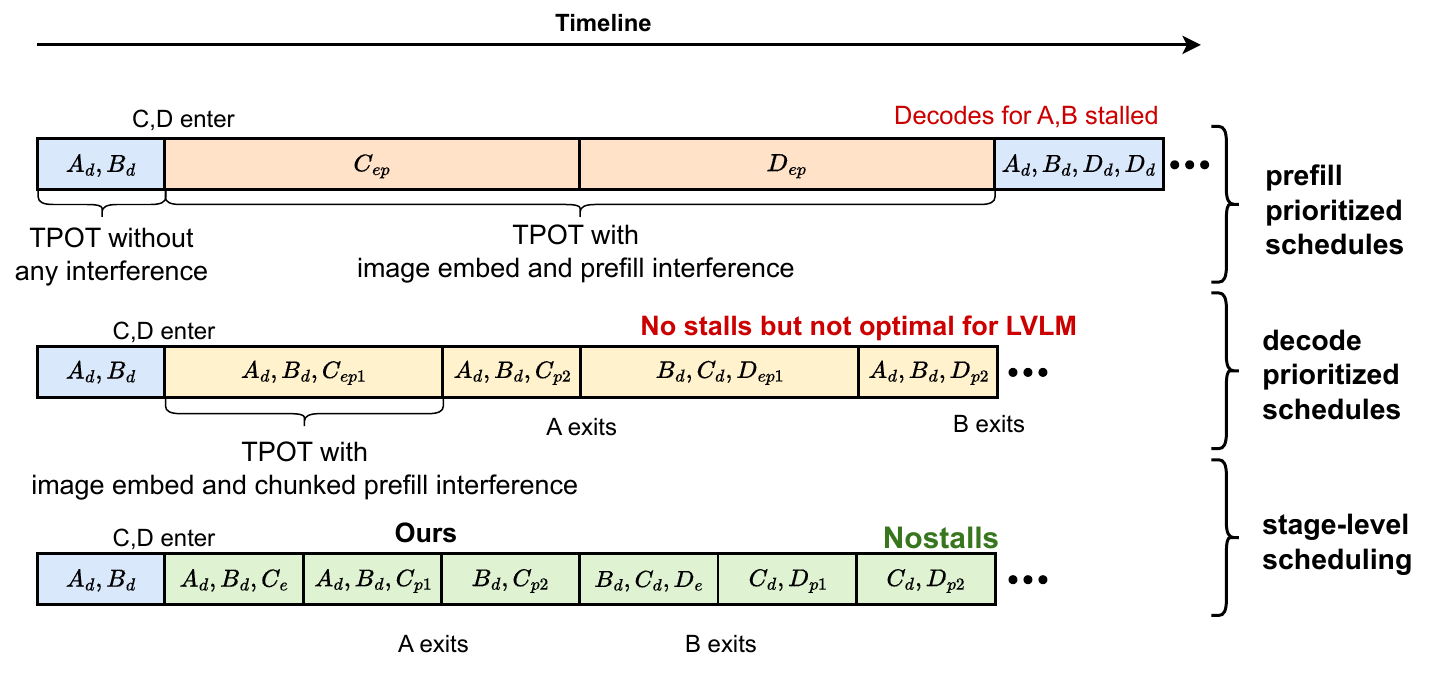}
  \caption{Comparison of the generation stall problem under different scheduling strategies. A, B, C, and D represent different requests. The subscript "e" denotes the image encode stage, "p" denotes the complete prefill stage, and "d" denotes the decode stage. "ep" represents the serial execution of encode and prefill, while "p1" and "p2" represent chunked prefill.}
  \label{fig:batch_schedule_policy}
\end{figure}

\textbf{Batching.} To improve GPU utilization, large language model serving systems employ batching techniques to process multiple requests simultaneously. A larger batch size helps amortize the cost of accessing model parameters across multiple requests~\cite{sarathi-serve, orca}.
Figure~\ref{fig:batchsize_analysis} shows how throughput changes with batch size. 
We observe that the encode stage reaches saturation at a batch size of around 6. 
The prefill stage achieves saturation even with a single request. 
The decode stage shows roughly linear throughput improvement with increasing batch size, saturating at around 512. 
Once the batch size exceeds the saturation point for each stage, increasing it further does not lead to additional throughput gains, while latency increases linearly.

\textbf{Takeaway-2:} Different stages achieve their maximum throughput at different batch sizes. Beyond that point, larger batch sizes lead to a linear increase in latency.

\textbf{Trading off Throughput and TBT Latency.} Batch processing enhances throughput but can lead to high tail TBT latency when decode tasks are included, a phenomenon known as \textit{generation stall}~\cite{sarathi-serve}.
Figure~\ref{fig:batch_schedule_policy} illustrates different scheduling strategies with four requests (A, B: decode stage; C, D: new arrivals) on a timeline. 
Prefill-prioritized scheduling increases throughput by prioritizing the prefill stage of new requests, but causes decode requests to stall, thereby increasing tail TBT latency.
In contrast, stall-free scheduling~\cite{sarathi-serve}, used by Sarathi Serve, prioritizes decode tasks to reduce tail TBT latency, though it slightly lowers throughput. This method batches both prefill and decode requests together, allowing continuous decoding as new requests arrive.
To mitigate the impact of new prefill requests, chunked prefill technology segments prompts for partial processing.  Additionally, stall-free scheduling also limits the maximum token count per batch to control execution time and ensure TBT latency guarantees. However, this approach is less effective for MLLMs, as it does not account for image encoding latency.

\textbf{Takeaway-3:} Existing scheduling strategies for language models struggle to precisely control execution time in multimodal inference, frequently resulting in violations of TBT SLOs.

\subsection{Difficulty in Disaggregation Method Selection}
\begin{figure}[t]
  \centering
  \includegraphics[width=\linewidth]{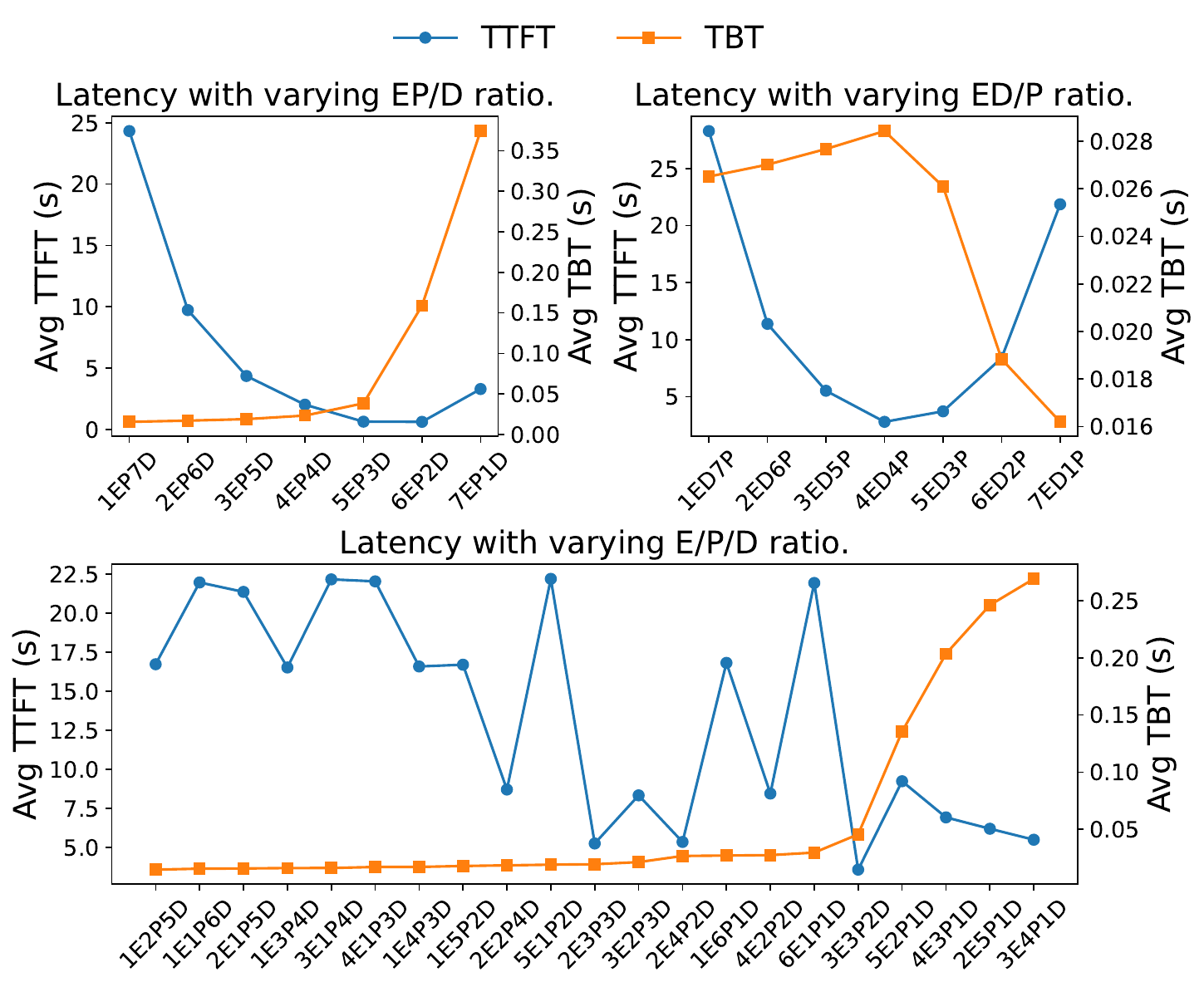}
  \caption{The effect of instance ratios on average TBT and TTFT under the three disaggregation methods.}
  \label{fig:instance_ratio}
\end{figure}

\begin{figure}[t]
  \centering
  \includegraphics[width=\linewidth]{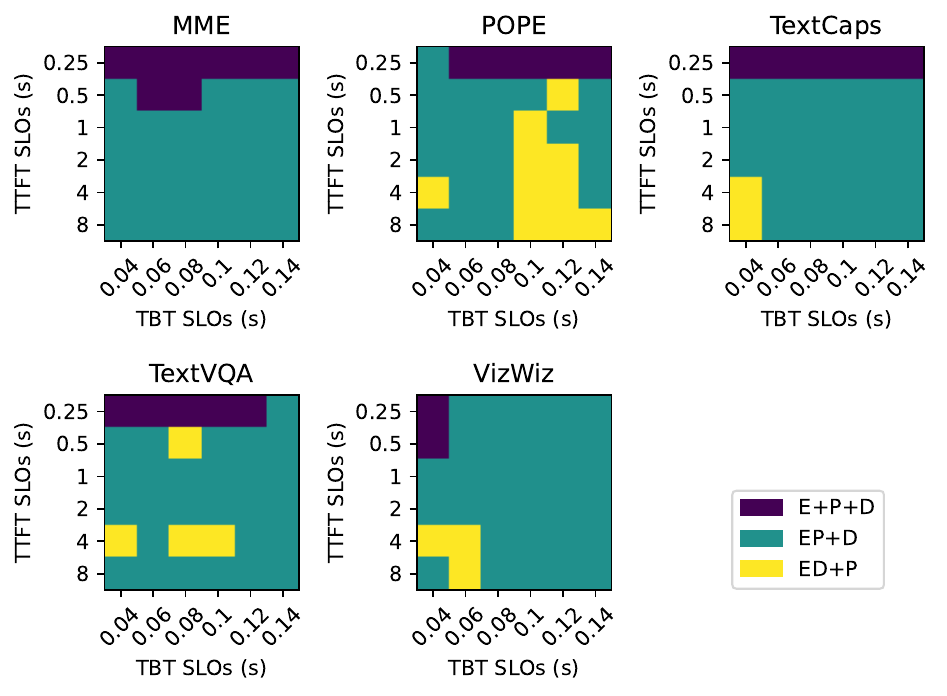}
  \caption{The effect of workload and SLOs on optimal disaggregation method selection. }
  \label{fig:disaggregation_methods}
\end{figure}

\textbf{Various Disaggregated Architures.} In multimodal inference, some possible existing disaggregation methods include: 
(1) \textit{E+P+D}: This design frees the encode instance from storing the KV cache and loading the language model, allowing for a larger image cache. This increases throughput by maximizing batch size without execution time constraints and improves cache hit rates for repeated images.
(2) \textit{EP+D}: Only one KV cache transfer is required, helping to reduce transfer overhead. And both the vision model and the language model can be processed in parallel.
(3) \textit{ED+P}: The vision model and language model can be processed in parallel, with the prefill stage remaining independent, which helps reduce TTFT. 

\textbf{Effect of Different Instance Ratios.} Figure~\ref{fig:instance_ratio} shows how different instance ratios affect performance using the TextCaps dataset (8 requests/sec).
In the \formatDisaggregationMethod{EP+D} architecture, a low number of \formatDisaggregationMethod{EP instances} (e.g., \formatDisaggregationMethod{1EP+7D}) leads to high computational loads and increased TTFT. Conversely, more \formatDisaggregationMethod{D} instances yield lower TBT, but as \formatDisaggregationMethod{EP} instances increase and \formatDisaggregationMethod{D} instances decrease, TBT rises. Specifically, at \formatDisaggregationMethod{7EP+1D}, the lack of \formatDisaggregationMethod{D} instances causes decode bottlenecks, increasing queuing delays and TTFT. 
In the \formatDisaggregationMethod{ED+P} disaggregated architecture, insufficient \formatInstanceType{ED} instances lead to bottlenecks in both the encode and decode stages, negatively impacting both TTFT and TBT. When \formatInstanceType{P} instances are insufficient, the prefill stage becomes the primary bottleneck, similarly resulting in increased TTFT.
For the fully disaggregated \formatDisaggregationMethod{E+P+D} architecture, results indicate a negative correlation between TBT and the number of \formatInstanceType{D} instances. Additionally, certain \formatInstanceType{E}-to-\formatInstanceType{P} instance ratios can significantly reduce TTFT for a given number of \formatInstanceType{D} instances.

\textbf{Takeaway-4}: Optimal instances ratios of different disaggregation methods depends on the requirements of the SLO.

\textbf{Effect of Different Disaggregation Methods.} 
Existing systems ~\cite{singh2024efficiently, qiu2025towards, guo2025rserveoverlappingencodingprefill} rely on a single, static disaggregation method that does not adapt to different scenarios, leading to suboptimal goodput performance. To investigate this, we conducted experiments using the LLaVA-Next-7B model, examining how different workload characteristics and SLO thresholds (TBT/TTFT) influence the selection of disaggregation methods. As shown in Figure~\ref{fig:disaggregation_methods}, the fully disaggregated \formatDisaggregationMethod{E+P+D} architecture shows significant advantages under stringent TTFT constraints. 
In contrast, \formatDisaggregationMethod{ED+P} and \formatDisaggregationMethod{EP+D} show superior performance in other scenarios. 

\textbf{Takeaway-5}: In multimodal large models, various disaggregation methods such as \formatDisaggregationMethod{E+P+D}, \formatDisaggregationMethod{EP+D}, and \formatDisaggregationMethod{ED+P} should be weighed against load characteristics and SLOs to achieve the maximization of resource utilization. 
\section{Design and Implementation}
\begin{figure*}[t]
  \centering
  \includegraphics[width=\linewidth]{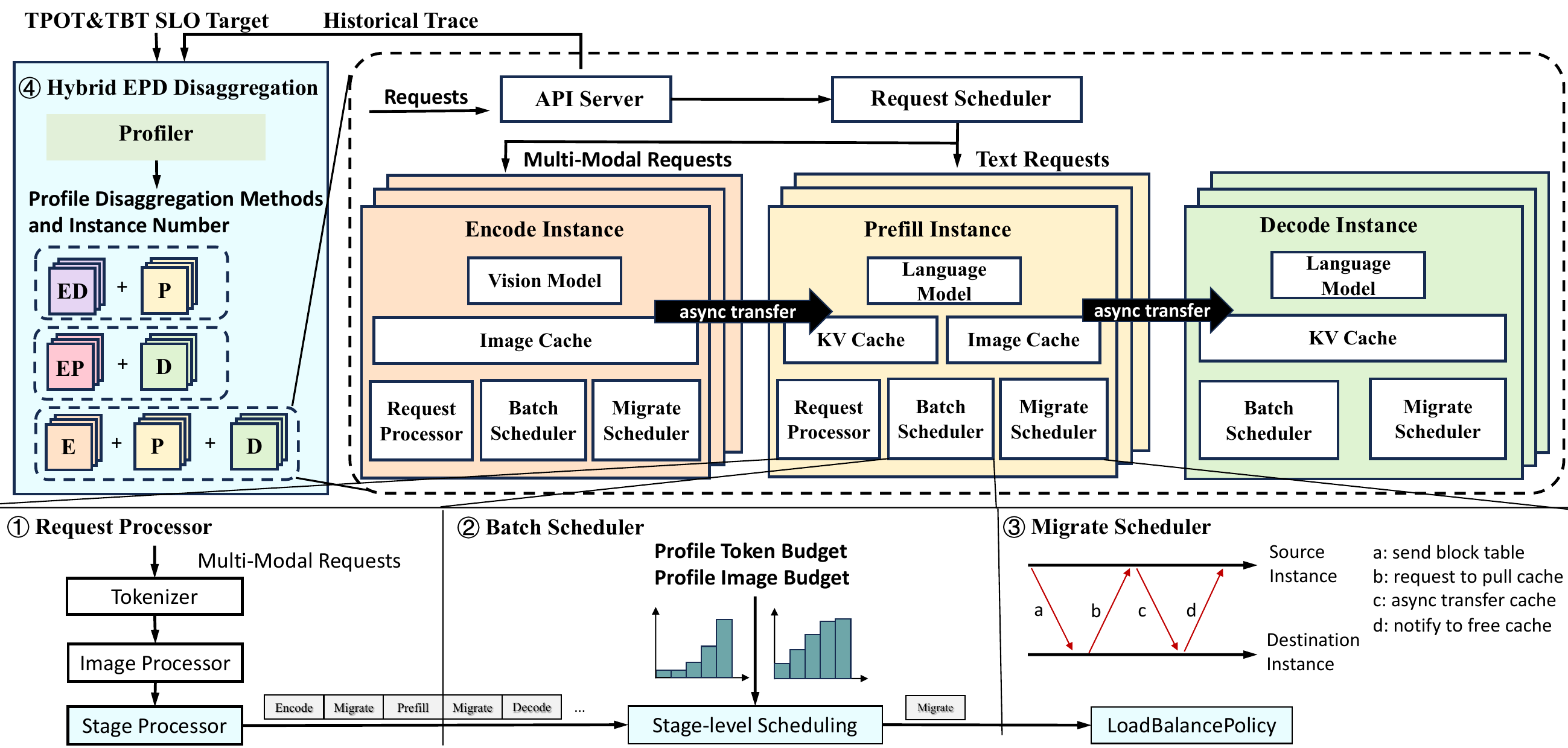}
  \caption{\formatSystemname{\systemname{}} system architecture.}
  \label{fig:system_design}
\end{figure*}

As illustrated in Figure~\ref{fig:system_design}, we present \formatSystemname{\systemname{}} to address the aforementioned challenges with its core component, Hybrid EPD Disaggregation (§~\ref{sec:hyrid_epd_disaggregation}), which dynamically selects the optimal disaggregation method and instance configurations.
The system receives client requests via the API Server, which forwards them to the Request Scheduler. Based on request types, the scheduler performs load balancing and dispatches them to the corresponding Encode or Prefill instances. Each type of instance loads specific model on demand and allocates its own cache space accordingly.
Within each instance, a Batch Scheduler (§~\ref{sec:batch_scheduler}) aggregates requests to improve processing parallel efficiency, while the Migrate Scheduler (§~\ref{sec:migrate_scheduler}) manages request migration between instances to achieve dynamic load balancing. Additionally, the Request Processor (§~\ref{sec:request_processor}) handles preprocessing of newly received requests.

\subsection{Request Processor}
\label{sec:request_processor}
To enable a stage-level scheduling strategy, we introduce a Request Processor that preprocess incoming requests. The Request Processor first performs tokenization and image processing before passing the requests to the Stage Processor, which then transforms it into a sequence of tasks (encode, prefill, decode, or migrate) and proactively prepares control parameters for these tasks. Finally, these tasks are dispatched to the Batch Scheduler for intra-instance scheduling. This design is motivated by the following considerations: By offloading part of the request scheduling computation to the Request Processor ahead of time, we reduce the CPU overhead during subsequent autoregressive inference. This preprocessing can be parallelized, mitigating the bottleneck caused by CPU-intensive tasks (e.g., generating control parameters for inference) and I/O-intensive operations (e.g., image processing). Additionally, this architecture enhances system extensibility, allowing for the easy integration of additional stages and token pruning optimizations.

\subsection{Batch Scheduler}
\label{sec:batch_scheduler}
\begin{algorithm}[t]
    \caption{Stage-level Batching. }
    \label{alg:batchpolicy}
    \renewcommand{\algorithmicrequire}{\textbf{Input:}}
    \renewcommand{\algorithmicensure}{\textbf{Output:}}
    \begin{algorithmic}[1]
        \REQUIRE Instance type $T$, maximum TTFT required for SLO $TTFT_{max}$, maximum TBT required for SLO $TBT_{max}$, the queue for waiting requests $Q_{waiting}$, the set for running requests $S_{running}$.
        \ENSURE batch of requests $B$. 
        \IF {$T$ in \{``E", ``EP", ``P"\} }
            \STATE $LATENCY_{max} \leftarrow \alpha * TTFT_{max}$
        \ELSIF{$T$ in \{``ED", ``EPD", ``D", ``PD"\}}
            \STATE $LATENCY_{max} \leftarrow TBT_{max}$
        \ENDIF
        \STATE Initialize $\tau_t, \tau_e\leftarrow $ search\_budget($LATENCY_{max}$)
        \STATE Initialize $n_t \leftarrow 0$, $n_e \leftarrow 0$, $B \leftarrow \emptyset$
        \FOR{$R$ in $S_{running}$}
            \IF{is\_decode\_stage($R$)}
                \STATE $n_t \leftarrow n_t + 1$; $B \leftarrow B \cup \{R\}$
            \ENDIF
        \ENDFOR
        \FOR{$R$ in $S_{running}$}
            \IF{is\_prefill\_stage($R$) and $n_t < \tau_t$}
                \STATE $n_t \leftarrow n_t + $ get\_chunked\_size($R$);
                \STATE $B \leftarrow B \cup \{R\}$
            \ENDIF
            \IF{is\_encode\_stage($R$) and $n_e < \tau_e$}
                \STATE $n_e \leftarrow n_e + $ get\_image\_number($R$);
                \STATE $B \leftarrow B \cup \{R\}$
            \ENDIF
        \ENDFOR
        \WHILE{$n_t < \tau_t$}
            \STATE $R \leftarrow$ get\_next\_text\_request($Q_{waiting}$)
            \STATE $n_t \leftarrow n_t + $ get\_chunked\_prefill\_size($R$);
            \STATE $B \leftarrow B \cup \{R\}$
        \ENDWHILE
        \WHILE{$n_e < \tau_e$}
            \STATE $R \leftarrow$ get\_next\_multi\_media\_request($Q_{waiting}$)
            \STATE $n_e \leftarrow n_e + $ get\_image\_number($R$);
            \STATE $B \leftarrow B \cup \{R\}$
        \ENDWHILE
        \STATE \textbf{Return} $B$
    \end{algorithmic}
\end{algorithm}

\textbf{Stage-level Scheduling.} We propose a unified scheduling strategy utilizing a Stage-level batching algorithm, which is designed for different types of instances and operates at the stage granularity in an iterative manner, enabling intra-instance request scheduling. 
As illustrated in Algorithm~\ref{alg:batchpolicy}, 
the Stage-level scheduling first sets the image budget for vision model and the token budget for language model based on the user-defined SLO. 
Specifically, for the \formatInstanceType{E}, \formatInstanceType{EP}, and \formatInstanceType{P} instances, we ensure the TTFT SLOs of the requests by limiting the execution time of each batch to $\alpha * TTFT_{max}$ (line 1-2), where $\alpha$ is a constant. In our scheduling strategy, we set it to 0.5, as each request must go through at least two iterations to output the first token: the first iteration completes the image encoding and the second one finalizes the prefill stage. For \formatInstanceType{ED}, \formatInstanceType{EPD}, \formatInstanceType{D}, \formatInstanceType{PD} instances, we limit the execution time of each batch to $TBT_{max}$ to meet the TBT SLOs (line 3-4). As the excution latency of a batch generally increases monotonically with the batch size (Figure~\ref{fig:batchsize_analysis}), we use binary search to profile the maximum encode batch size (also called image budget) and maximun token batch size (also called token budget) during system initialization (line 5). This design ensures the maximization of system throughput while adhering to the TTFT and TBT SLOs defined by user requirements.
As shown in Figure ~\ref{fig:batch_schedule_policy}, after system initialization, each iteration first includes all ongoing decode requests into the current batch (line 7-9). 
Then, we check if there are any partially computed chunked prefill tasks or encode tasks and include them in the batch if available (line 10-16). 
Finally, new requests are added to fill $\tau_t$ and $\tau_e$, in order to maximize throughput (line 17-24). 


\textbf{Dual-stream Parallelism.} Inspired by NanoFlow~\cite{zhu2024nanoflowoptimallargelanguage}, we employ dual CUDA streams to achieve kernel-level parallelism between the vision model and the language model. Specifically, one vision stream is dedicated to image encoding task, the other language stream is allocated for prefill and decode tasks. 
In sequential execution, low compute utilization during the decode stage or low memory utilization during prefill stage results in underutilized GPU resources; instead, parallel multi-stream execution enhances throughput for both the language model and the vision model, thereby optimizing GPU resource utilization.

\subsection{Migrate Scheduler}
\label{sec:migrate_scheduler}
The Migrate Scheduler is responsible for request migration using a pull-based model to prevent cache overflows at the receiving instance, mitigating issues such as KV cache or image cache exhaustion.
The migration process consists of four steps. Firstly, the Migrate Scheduler at the source instance sends the request’s control information, including KV and image cache page tables, to the target instance, which enqueues the request at the head of $Q_{waiting}$. Secondly, when the request is scheduled, the target instance creates new page tables and requests the cache blocks from the source.
Thirdly, the source instance asynchronously transfers the KV and image caches to the target instance. Finally, after the migration is complete, the target instance notifies the source instance to release the resources associated with the request.
Additionally, the Migrate Scheduler manages load balancing across multiple target instances by employing a default round-robin strategy, which also supports customizable options such as random selection, least-load priority, and prioritization of instances with higher prefix cache hit rates.

\subsection{Hybrid EPD Disaggregation}
\label{sec:hyrid_epd_disaggregation}
To achieve an efficient disaggregated inference system, we design a hybrid EPD disaggregation architecture, as illustrated in Figure~\ref{fig:system_design}. Built on a general-purpose inference engine, the system flexibly generates specialized instances (E, P, D, EP, ED) by configuring parameters such as the number of KV cache blocks and image cache blocks. Each instance internally adopts a unified task batching and scheduling strategy, as detailed in Algorithm~\ref{alg:batchpolicy}.

\begin{algorithm}[t]
    \caption{Hybrid EPD Disaggregation Profiler.}
    \label{alg:disaggregation_profiler}
    \renewcommand{\algorithmicrequire}{\textbf{Input:}}
    \renewcommand{\algorithmicensure}{\textbf{Output:}}
    \begin{algorithmic}[1]
        \REQUIRE Historical trace $H$ and its number of requests$N_r$, number of instances $N$, GPU memory capacity $C$, model weight memory cost $C_{model}$, TTFT SLO $TTFT_{max}$, TBT SLO $TBT_{max}$
        \ENSURE a mapping from instance type to number S.
        \STATE Initialize workload $W_e \leftarrow 0$,$W_p \leftarrow 0$,$W_d \leftarrow 0$
        \FOR{each request $R$ in $H$}
            \STATE $W_e \leftarrow W_e + $ get\_visual\_token\_number($R$)
            \STATE $W_p \leftarrow W_p + $ get\_prompt\_and\_visual\_token\_number($R$)
            \STATE $W_d \leftarrow W_d + $ get\_decode\_token\_number($R$)
        \ENDFOR
        \STATE $\tau_e \leftarrow$ search\_image\_budget($\alpha * TTFT_{max}$)
        \STATE $\tau_p \leftarrow$ search\_token\_budget($\beta * TTFT_{max}$)
        \STATE $\tau_d \leftarrow$ search\_token\_budget($TBT_{max}$)
        \STATE $\tau_d \leftarrow$ min($\tau_d, \gamma * \frac{(C-C_{model})*N_r}{W_p + W_d}$)
        \STATE $tp_e, tp_p, tp_d \leftarrow$ profile\_throughput($\tau_e$, $\tau_p$, $\tau_d$)
        \STATE $t_e \leftarrow \frac{W_e}{tp_e}$; $t_p \leftarrow \frac{W_p}{tp_p}$; $t_d \leftarrow \frac{W_d}{tp_d}$
        \STATE $N_e, N_p, N_d \leftarrow$ partition($N, t_e, t_p, t_d$)
        \STATE Initialize \( best\_goodput \leftarrow 0 \)
        \STATE Let \( E\_P\_D \leftarrow \{ ``E": N_e, ``P": N_p, ``D": N_d \} \)
        \STATE Let \( EP\_D \leftarrow \{ ``EP": N_e + N_p, ``D": N_d \} \) 
        \STATE Let \( ED\_P \leftarrow \{ ``ED": N_e + N_d, ``P": N_p \} \) 
        \FOR{each $method$ in $\{E\_P\_D, EP\_D, ED\_P\}$}
            \STATE start\_inference\_server($method$)
            \STATE $goodput \leftarrow$ replay\_trace($H$)
            \IF{$goodput > best\_goodput$}
                \STATE $best\_goodput\leftarrow goodput$
                \STATE $best\_method\leftarrow method$
            \ENDIF
        \ENDFOR
        \STATE \textbf{Return} $best\_method$
    \end{algorithmic}
\end{algorithm}
At the deployment level, instances are composed into different disaggregation methods based on task functionalities, such as \formatDisaggregationMethod{EP+D}, \formatDisaggregationMethod{ED+P}, and \formatDisaggregationMethod{E+P+D}, to support diverse inference workflows. 
Initially, the system runs with the default disaggregation method, with general EPD instance, while collecting a period of request trace that include image resolution, prompt length, decode length, and timestamps. Under the assumption of a stable workload, we design a heuristic search algorithm (Algorithm~\ref{alg:disaggregation_profiler}) to perform workload analysis and request replay on historical data to find the optimal disaggregation method.
The algorithm begins by calculating the total number of tokens for all request stages (line 1-5).
By calculating the maximum batch size for each batch under full load, it then estimates the throughput (tokens per second) of E, P, and D instances to calculate the execution time required for these workloads. Specificallly, we first binary search the image budget and token budget for different instances based on the SLOs of TTFT and TBT (line 6-8), where $\alpha$ and $\beta$ are empirical parameters set to 0.5. Using historical trace, we then estimate the cache required for each request to calculate the maximum number of concurrent requests that can be executed in the \formatInstanceType{D} instance (line 9), with memory utilization used for cache $\gamma$ set to 0.9. The minimum value of the token budge and the maximum number of concurrent requests determines the maximum batch size per iteration under full load (line 10). Next, we obtain the throughput under full load for different instances through profiling (line 11).
We then partition the entire cluster based on the ratio of the three execution times, rounding the values to ensure that there is at least one instance is allocated for each stage(line 12).
Finally, we test these three disaggregation methods by replaying the historical request trace, measuring the goodput, and determining the optimal disaggregation method (line 13-22).


\textbf{Complexity.}
Since both \formatDisaggregationMethod{EP+D} and \formatDisaggregationMethod{ED+P} have $C_{N-1}^1$ combination ways, and \formatDisaggregationMethod{E+P+D} has $C_{N-1}^2$ combination ways, the total search space for our heuristic search algorithm is given by $2C_{N-1}^1+C_{N-1}^2$, where $N$ is the total number of instances.
In comparison to brute force search, our heuristic search is theoretically $\frac{3}{2C_{N-1}^1+C_{N-1}^2}$ times faster.
In actual measurements, the time required to estimate throughput is approximately one minute. Since we can reuse the profiling results during system startup from the batch scheduler, this time is negligible. For request replay, we only sample requests for a few minutes, so the total execution time of the algorithm is approximately serveral minutes.

\textbf{Redeploy.} The disaggregation method and number of instances are fine-tuned for a particular workload pattern.
However, this setup may become inefficient if the workload evolves overtime. To address this, \formatSystemname{\systemname{}} performs periodic adjustments. 
We regularly analyze historical request trace and whenever a noticeable shift in the workload pattern is detected, \formatSystemname{\systemname{}} triggers a rerun of the algorithm using the latest historical trace. 
This process is highly efficient with the algorithm completing in several minutes and reloading model weights taking only a few minutes – much faster than the hourly intervals typically observed in real-world workload fluctuations.
Additionally, we may consider reusing existing instances to speed up the redeployment process and reduce the overhead of model reloading, which is part of future work.

\subsection{Implementation}
\label{sec:implementation}
\formatSystemname{\systemname{}} is an efficient online inference system for MLLMs, comprising approximately 10K lines of Python code and 3K lines of C++ and CUDA code. The frontend and inference engine are implemented in Python, while the data transmission modules and model computation kernels are implemented in C++ and CUDA.
It adopts a RESTful API frontend that connects to multiple parallel inference engine instances, compatible with OpenAI-style APIs.
We build the multi-instance inference system using Ray~\cite{ray}, where each instance is implemented as a Ray actor. Data transfer for the image cache and KV cache is enabled through CUDA IPC memory handles~\cite{ipc} and NCCL~\cite{nccl}. We utilize the FlashAttention~\cite{dao2023flashattention2fasterattentionbetter} and FlashInfer~\cite{flashinfer} kernels to implement page attention~\cite{vllm}.
Additionally, to facilitate request migration and cache reuse between requests, we employ a paged management approach for image caching like kv cache. 

\section{Evaluation}

\subsection{Experiments Setup}
\begin{figure}[t]
  \centering
  \includegraphics[width=\linewidth]{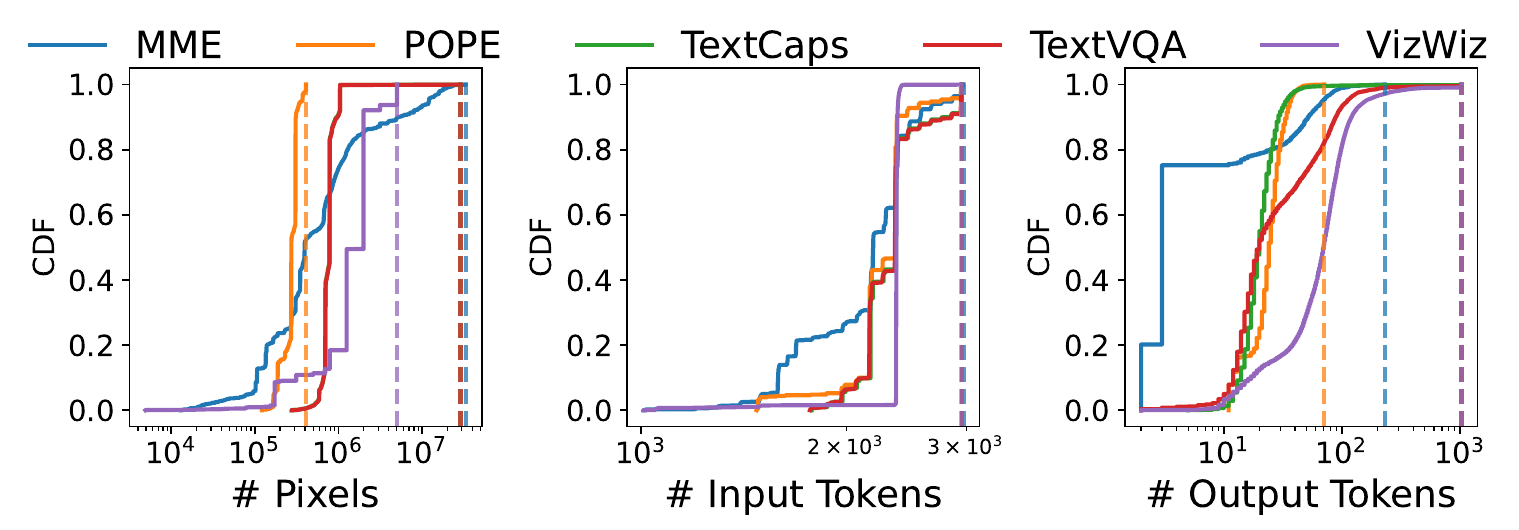}
  \caption{The workload of the LLaVA-NeXT-7B model across different datasets.}
  \label{fig:dataset_workload_analysis}
\end{figure}

\begin{figure*}[t]
  \centering
  \includegraphics[width=\linewidth]{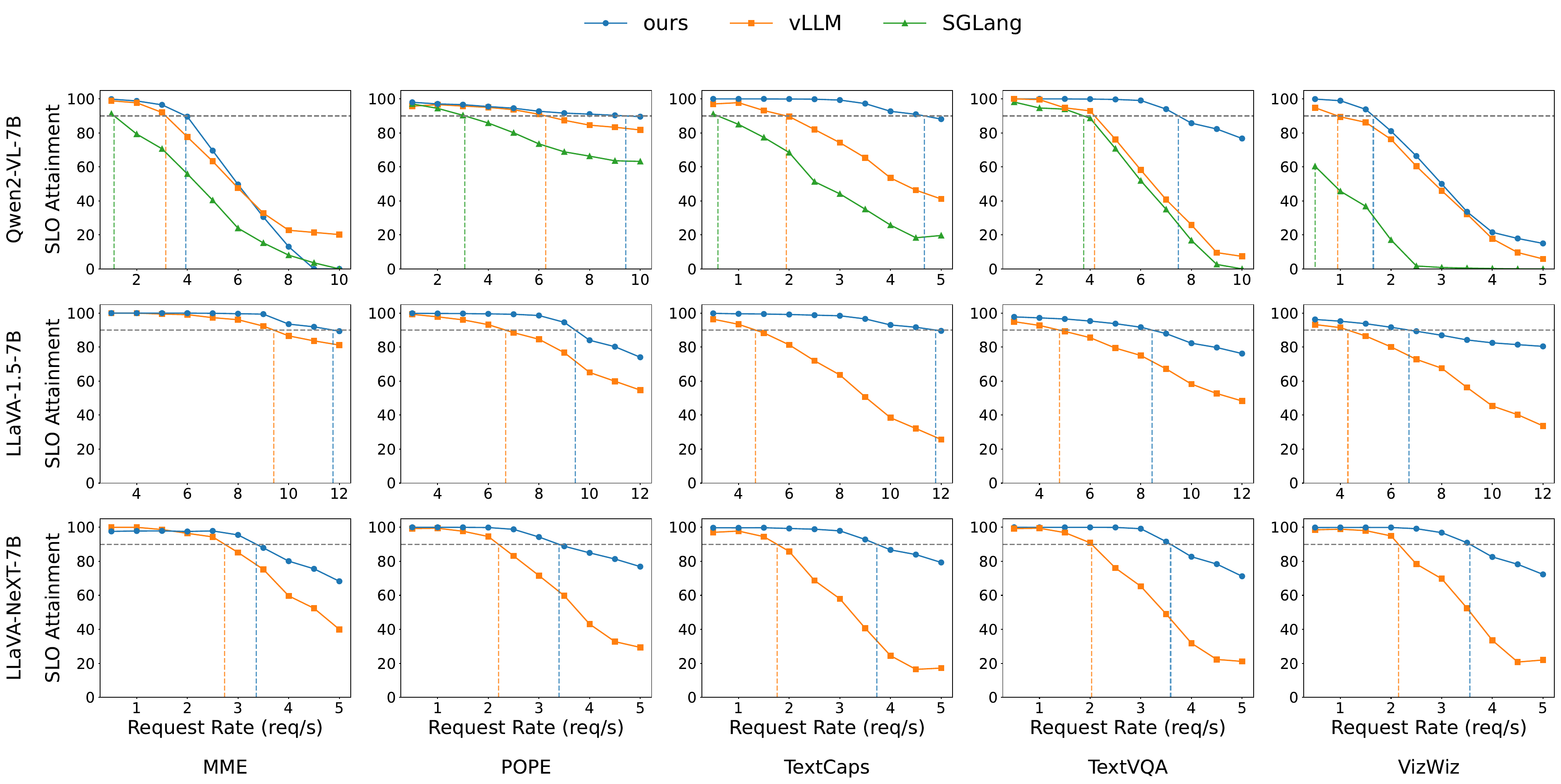}
  \caption{Comparison of SLO attainment, the Request Rate is the average load per-GPU.}
  \label{fig:main_evaluation}
\end{figure*}
\textbf{Cluster Testbed}. We deploy \formatSystemname{\systemname{}} on 4 servers, each equipped with 8 NVIDIA H20 GPUs (141 GB), 2$\times$200Gbps InfiniBand NICs, connected with NVLink, 96-CPU cores, and 2TB of RAM. \\
\textbf{Baselines}. 
We compare our system with 
vLLM (version 0.11.0)~\cite{vllm}, 
and SGLang (version 0.5.3)~\cite{sglang}, both of which are fast serving framework for large language models and vision language models. \\
\textbf{Models}. We test various models to demonstrate the generalization ability of our system. Specifically, we evaluate LLaVA-1.5-7B ~\cite{llava1.5}, LLaVA-NeXT-7B~\cite{llavanext}, and Qwen2-VL-7B ~\cite{qwenvl}. \\
The number of tokens generated for the same image varies across these models, which in turn impacts the request load. LLaVA-1.5 encodes each image into 576 tokens, while LLaVA-NeXT and Qwen2-VL employ different methods to adjust the number of tokens based on the image resolution. \\
\textbf{Workloads}. 
We use diverse datasets to encompass various application scenarios:
TextCaps~\cite{textcap}, 
POPE~\cite{pope}, 
MME~\cite{mme}, 
TextVQA~\cite{textvqa}, 
VizWiz~\cite{vizwizvqa}. 
We extract one-tenth of the dataset as historical traces to profile the disaggregation method, and then use the remaining data for performance evaluation.
To ensure consistent load across different baselines, we first perform inference on the datasets and record the number of tokens generated during decoding. 
For subsequent tests, we set the max tokens parameter and the ignore eos parameter to fix the number of output tokens.
Different datasets have varying workloads. The results obtained from LLaVA-NeXT-7B inference are shown in Figure~\ref{fig:dataset_workload_analysis}. To ensure realistic workload distribution, we adopt the inter-arrival timestamps of requests from the production inference trace~\cite{qin2024mooncake}, scaling the timestamps to achieve different request rates (requests per second). \\
\textbf{Details}. The testing environment uses CUDA version 12.4. The KV cache block size is 16, while image cache block size is 576. All models, along with intermediate variables, KV cache, and image cache, are set to fp16 precision. vLLM runs in eager mode. All inference engines employ the same chat template and both prefix cache and CUDA graph afeatures are disabled.

\subsection{SLO Attainment Evaluation}
In this section, We compare the SLO attainment of different methods at various request rates across different workloads. The SLO settings are detailed in Appendix~\ref{sec:slo_settings}. 
For the baseline, we deploy 32 instances and use round-robin scheduling.

The experimental results are presented in Figure~\ref{fig:main_evaluation}. The gray dashed line indicates the 90\% SLO attainment threshold, while the vertical dashed line marks the request rate at which each inference system achieves 90\% SLO attainment, measured as goodput.
{It is evident that our system consistently achieves significantly higher goodput than the other systems. Specifically, for the Qwen2-VL-7B model, our system outperforms SGlang by \minGoodputImprovementQwentwoVLsevenBSGLangAcrossDatasets{}x to \maxGoodputImprovementQwentwoVLsevenBSGLangAcrossDatasets{}x in goodput. For the LLaVA-1.5-7B and LLaVA-NeXT-7B models, we achieve goodput improvements of  \minGoodputImprovementLLaVAonepointfivesevenBvLLMAcrossDatasets{}x–\maxGoodputImprovementLLaVAonepointfivesevenBvLLMAcrossDatasets{}x and \minGoodputImprovementLLaVANeXTsevenBvLLMAcrossDatasets{}x-\maxGoodputImprovementLLaVANeXTsevenBvLLMAcrossDatasets{}x compared to vLLM, respectively. This performance gain is attributed to our Hybrid EPD Disaggregation architecture, which effectively selects appropriate disaggregation methods and reduces interference between different execution stages, thereby lowering TBT for requests and increasing the SLO attainment rate. However, the improvement is smallest for the MME dataset, as the requests in this dataset require a minimal number of tokens to be output, limiting the optimization effect of our scheduling method on TBT. Nevertheless, the hybrid disaggregation architecture and multi-stream parallelism still contribute to some throughput optimization.

\subsection{Ablaion Study}
\begin{figure}[t]
  \centering
  \includegraphics[width=0.95\linewidth]{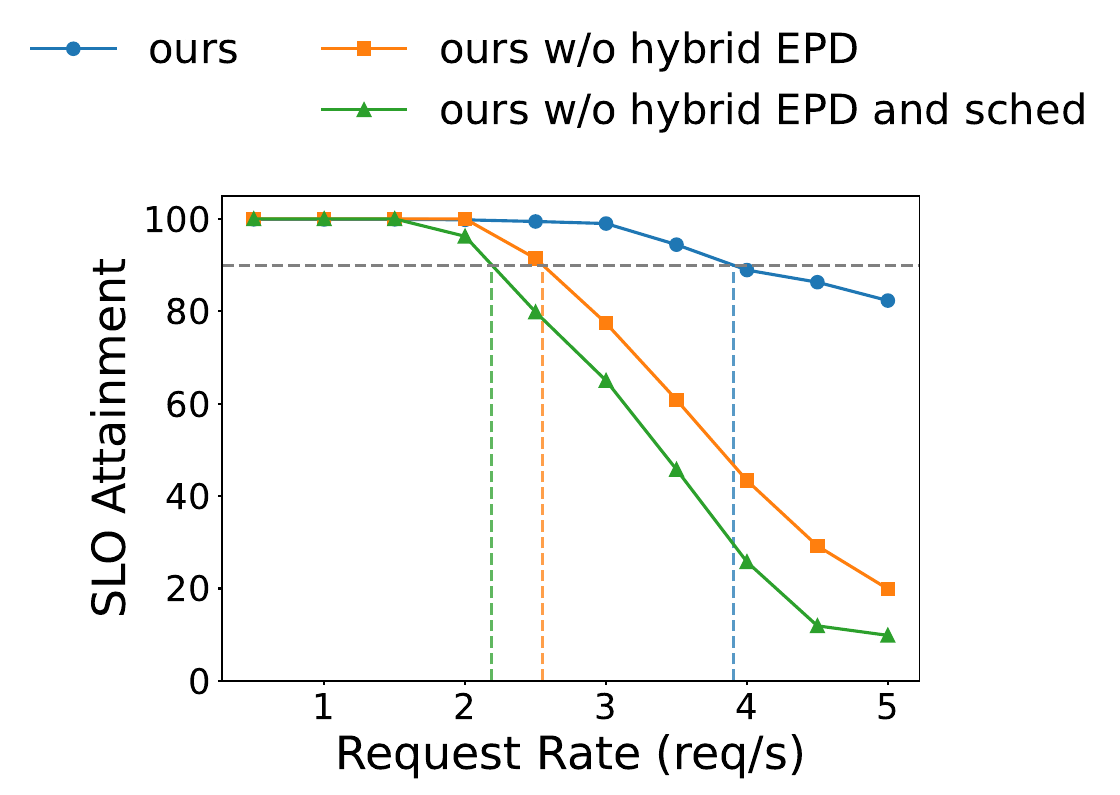}
  \caption{Ablation study of hybrid EPD disaggregation and batch schedule policy.}
  \label{fig:abligation}
\end{figure}

To validate the effectiveness of our scheduling strategy (Algorithm~\ref{alg:batchpolicy}) and Hybrid EPD Disaggregation (§~\ref{sec:hyrid_epd_disaggregation}) in controlling TBT and improving the SLO achievement rates, we conducte a comparative experiment on the TextCaps dataset using the LLaVA-Next-7B model.
As shown in Figure~\ref{fig:abligation}, we first disable the Hybrid EPD Disaggregation method and use a \formatDisaggregationMethod{E+P+D} disaggregation architecture with the cluster instances evenly partitioned into three parts. In this configuration, we observed a drop in goodput from \goodputours{} to \goodputourswohybridEPD{} req/s, highlighting that an appropriately selected disaggregation strategy can significantly improve resource utilization and SLO satisfaction.
Subsequently, we disable the stage-level scheduling policy. This further reduction in goodput, from \goodputourswohybridEPD{} to \goodputourswohybridEPDandsched{} req/s, demonstrates that our stage-level scheduling strategy effectively limits the execution time of each batch, leading to tighter adherence to TBT.

\subsection{Latency Breakdown}

\begin{figure}[t]
  \centering
  \includegraphics[width=\linewidth]{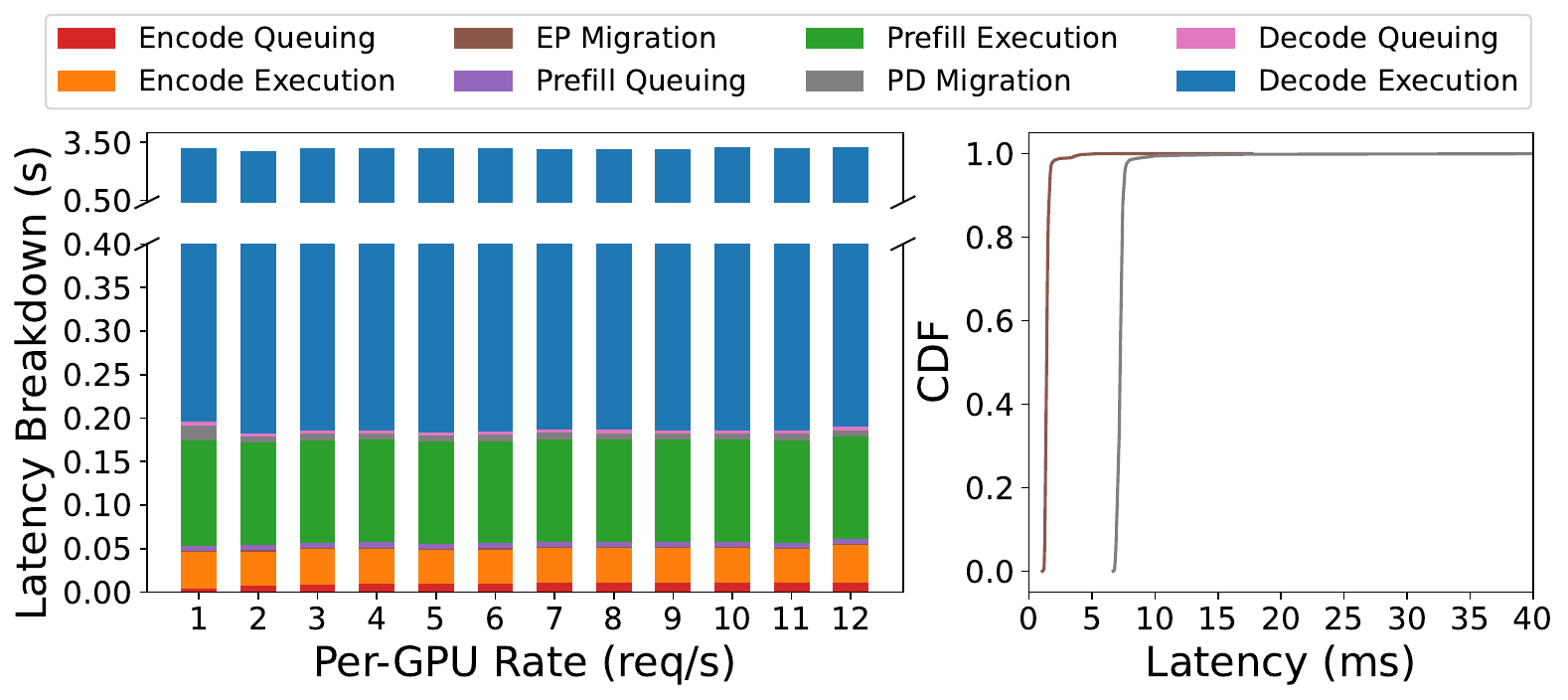}
  \caption{Latency breakdown when serving llava1.5-7B on textcaps dataset.}
  \label{fig:latency_breakdown}
\end{figure}

To better understand the performance characteristics of \formatSystemname{\systemname{}} and identify potential system bottlenecks, we conduct a detailed latency breakdown analysis of service requests. We divide the lifecycle of each request into multiple stages: encode queueing, encode execution, EP migration, prefill queueing, prefill execution, PD migration, decode queueing, and decode execution. We measure the average latency for each stage using the LLaVA-1.5-7B model on the TextCaps dataset with a \formatDisaggregationMethod{1E3P4D} disaggregation configuration. The latency distribution across stages is shown in Figure~\ref{fig:latency_breakdown}, indicating that the majority of the request latency is concentrated in the decode stage, followed by the prefill and encode stages.
The overhead introduced by image cache and KV cache migration is minimal, accounting for less than 1\% of the total latency. Specifically, 95\% of image cache migration complete within 2ms, and 95\% of KV cache migration complete within 8ms — both typically shorter than the execution time of a single decode batch. Therefore, their impact on TBT can be considered negligible.

\subsection{Dual-stream Parallelism}
\label{sec:dualstream_experiment}

\begin{figure}[t]
  \centering
  \includegraphics[width=\linewidth]{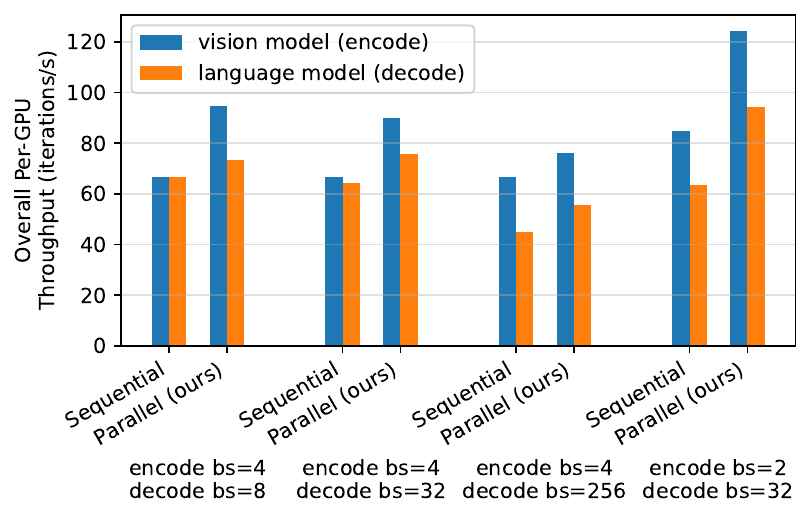}
  \caption{Overall per-GPU throughput when executing LLaVA-1.5-7B's vision model (encode) and language model (decode) sequential or parallel under different batch size. The KV cache length of decode is 1024.}
  \label{fig:multistream_concurrency_throughput}
\end{figure}

In this section, we investigate the impact of dual-stream parallelism on throughput during encode and decode stages using LLaVA-1.5-7B.
As shown in Figure~\ref{fig:multistream_concurrency_throughput}, the dual-stream parallelism significantly improves throughput of both the vision model and the language model compared to sequential execution across various batch sizes. This improvement arises from the simultaneous processing of the encode and decode stages in separate CUDA streams, which optimizes resource utilization and minimizes idle time. In contrast, the \textit{Sequential} executes the stages in a round-robin manner, with each stage taking 50\% share of the execution time. This approach inherently limits throughput, as it forces the system to wait for one stage to complete before starting the next, leading to increased idle periods and reduced overall efficiency. 

\section{Related Work}

\textbf{MLLM serving systems.}
Several works have focused on optimizing MLLM serving systems. Singh \textit{et al.}~\cite{singh2024efficiently} proposed EPD disaggregation to optimize resource allocation. 
Qiu \textit{et al.}~\cite{qiu2025towards} conducted a comprehensive performance analysis of MLLM serving characteristics and proposed a decoupled architecture concept with separate Image Nodes and Text Nodes. Zhong\textit{et al.}~\cite{guo2025rserveoverlappingencodingprefill} utilizes EPD disaggregation and orchestrates both intra- and inter-reauest pipelines to minimize latency and optimize paralelism. However, these works only consider the methods where the vision model and language model are either disaggregated or co-located but executed sequentially, overlooking the potential GPU efficiency gains from co-locating and executing them in multi-stream parallel. We comprehensively assess both decoupled and co-located multi-stream parallel strategies to identify the optimal model disaggregation method, including designs within the subset of E+P+D disaggregation methods. Our findings indicate that these approaches are not optimal in many scenarios. Notably, we did not directly evaluate these specific methods, as they are not open-sourced or implemented.
On the other hand, works like Inf-MLLM~\cite{ning2024inf} and Elastic Cache~\cite{liu2024efficient} reduce computational overhead or memory footprint through KV cache approximate reuse or pruning. These model-level approaches trade off model performance to reduce performance overhead, which is orthogonal to our system-level and scheduling-level optimizations that unchanged model performance. 
The widely used SOTA LLM inference frameworks, including SGLang~\cite{sglang}, vLLM~\cite{vllm}, and TGI~\cite{text-generation-inference}, also extended themselves to support MLLM model inference. However, all of them mix Encode, Prefill, and Decode within the same instances, leading to potential interference among the three stages, which can impact SLO guarantees adversely. 

\textbf{LLM scheduling optimization.}
Numerous works have explored optimizations for LLM scheduling.
Early works~\cite{orca, vllm} proposed continuous batching to improve GPU utilization. However, the co-located sequential execution of prefill and decode stages create a trade-off between TTFT and TBT. Sarathi-Serve introduced chunked-prefill~\cite{sarathi-serve}, attempting to balance TTFT and TBT by splitting long prefill stages into chunks and interleaving them with decode tasks. Nonetheless, this approach inherently sacrifices TTFT for TBT without fundamentally eliminating prefill-decode interference. DistServe~\cite{distserve} proposed a decoupled architecture that isolates prefill and decode stages on separate GPUs to avoid interference entirely, but introduces challenges of reduced GPU utilization and KV cache transfer overhead. While these works are not directly applicable to MLLM inference, their decoupling philosophy inspired subsequent multimodal research. Concurrently, works on KV cache reuse~\cite{qin2024mooncake, hu2024memserve}, KV cache migration optimization ~\cite{patel2024splitwiseefficientgenerativellm, jin2024p}, and placement dynamic adjustment~\cite{hu2024inferenceinterferencedisaggregatellm, patel2024splitwiseefficientgenerativellm} are orthogonal to our approach.
\section{Conclusion}
This paper proposes \texttt{\systemname}, an efficient system architecture for multimodal large model inference tasks. To address issues such as coupling during image-text processing and inflexible scheduling in existing systems, \texttt{\systemname{}} introduces the Hybrid Encode-Prefill-Decode Disaggregation design, enabling efficient scheduling and utilization of resources across different stages. \texttt{\systemname{}} supports stage-level batch processing and parallel execution mechanisms, improving the system's throughput. Experimental results demonstrate that the system can effectively handle diverse multimodal inference workloads while ensuring service quality, showcasing strong generality and performance advantages.

\bibliography{reference}
\bibliographystyle{mlsys2025}

\clearpage
\appendix
\section{SLO Settings}
\label{sec:slo_settings}
\begin{table}[h]
\caption{SLO settings under different workloads.}
\centering
\setlength{\tabcolsep}{2.9mm}{
\begin{tabular}{llrr}
\toprule
 Model         & Dataset   &   TTFT (s) &   TBT (s) \\
\midrule
 LLaVA-1.5-7B  & VizWiz    &       4    &       0.08 \\
 LLaVA-1.5-7B  & TextVQA   &       4    &       0.08 \\
 LLaVA-1.5-7B  & MME       &       4    &       0.08 \\
 LLaVA-1.5-7B  & POPE      &       4    &       0.08 \\
 LLaVA-1.5-7B  & TextCaps  &       4    &       0.08 \\
 LLaVA-NeXT-7B & VizWiz    &       8    &       0.60 \\
 LLaVA-NeXT-7B & TextVQA   &       8    &       0.60 \\
 LLaVA-NeXT-7B & MME       &       8    &       0.60 \\
 LLaVA-NeXT-7B & POPE      &       8    &       0.30 \\
 LLaVA-NeXT-7B & TextCaps  &       8    &       0.30 \\
 Qwen2-VL-7B   & VizWiz    &       8    &       0.60 \\
 Qwen2-VL-7B   & TextVQA   &       8    &       0.60 \\
 Qwen2-VL-7B   & MME       &       8    &       0.60 \\
 Qwen2-VL-7B   & POPE      &       8    &       0.10 \\
 Qwen2-VL-7B   & TextCaps  &       8    &       0.10 \\
\bottomrule
\end{tabular}
}
\label{tab:slo_settings}
\end{table}

As different models encode the same image into varying numbers of tokens, this leads to different prefill and decode workloads, which in turn affect the SLO configuration. Therefore, we define separate SLO thresholds for each model and dataset, as detailed in Table~\ref{tab:slo_settings}.
\section{Computation and Memory Access in MLLMs}
\label{sec:latency_model}


Table~\ref{tab:latency_model}, shows the symbol definitions. To simplify the model, we assume that all requests have the same prompt, and each request corresponds to a single image, with the image having the same token count. 
Different $H, M$ should been applied according to different model (vision or language).

In modern serving systems~\cite{vllm, sglang}, some operations like softmax and layernorm are usually fused with matrix multiplication kernels for efficiency~\cite{dao2023flashattention2fasterattentionbetter, flashinfer}. Therefore, attention and matrix multiplication in both feed-forward layers and attention layers dominate the overall computation and memory access, and our analysis focuses on them.

For the encode stage, the batch has $BT$ tokens. For QKVO projection in the attention layer, each matrix multiplication has one multiply and one add operation, so the arithmetic intensity is:
\begin{equation}
C^{qkvoproj}_{encode}=4 * BTH * 2 * H = 8BTH^2
\end{equation}
Since it is necessary to read the model parameters and inputs, and write the outputs, the memory access amount is:
\begin{equation}
M^{qkvoproj}_{encode} = 4(H^2+BTH+BTH) = 8BTH+4H^2
\end{equation}
For the feed forward layer, there are up projection and down projection, we assume intermediate size is $4H$. The arithmetic intensity is:
\begin{equation}
C^{ffn}_{encode}=2 * BTH * 2 * 4H = 16BTH
\end{equation}
The memory access amount is:
\begin{equation}
M^{ffn}_{encode}=2(4H^2 + BTH+BT4H) = 10BTH+8H^2
\end{equation}
The fused attention kernel involves multiplying the query with the key and multiplying the value, and we ignore softmax. The arithmetic intensity is approximately: 
\begin{equation}
C^{attention}_{encode} = 2 * (B*M*T^2 * 2*\frac{H}{M})=4BT^2H
\end{equation}
The reading of the query, key, and value, as well as the reading and writing of the attention scores and output, involve memory access. The amount is:
\begin{equation}
M^{attention}_{encode} = 4BTH+2BT^2M
\end{equation}
The prefill stage and decode stage are similar to the encode stage, except that the token count is $BS$ and $B$, respectively. Additionally, the decode stage needs to read the previous KV cache. Based on the above calculations, we summarize the computational and memory access requirements at different stages in Table~\ref{tab:latency_model}.


Through this analysis, we can roughly determine whether an operator is a computational bottleneck or a memory bottleneck, which in turn guides our scheduling design.

\end{document}